\documentclass[aps,pra,preprint,showpacs,amsmath,amssymb]{revtex4}
\input{epsf}

\usepackage{graphicx}
\usepackage{array}
\usepackage{bigstrut}
\usepackage{longtable}
\usepackage{rotating,booktabs}
\usepackage{booktabs,threeparttable}
\usepackage{bm}

\begin{document}

\title{Computational investigation of static multipole polarizabilities and sum rules for ground-state hydrogen-like ions}

\author{Li-Yan Tang$^{1}$, Yong-Hui Zhang$^{1,2}$, Xian-Zhou Zhang$^{2}$, Jun Jiang$^{3}$ and J. Mitroy$^{3}$}

\affiliation {$^1$State Key Laboratory of Magnetic Resonance and
Atomic and Molecular Physics, Wuhan Institute of Physics and
Mathematics, Chinese Academy of Sciences, Wuhan 430071, P. R. China}

\affiliation {$^{2}$Department of Physics, Henan Normal University,
XinXiang 453007, P. R. China}

\affiliation {$^{3}$School of Engineering, Charles Darwin
University, Darwin NT 0909, Australia}

\date{\today}

\begin{abstract}
High precision multipole polarizabilities, $\alpha_{\ell}$ for $\ell \le 4$
of the $1s$ ground state of the hydrogen isoelectronic series are obtained
from the Dirac equation using the B-spline method
with Notre Dame boundary conditions. Compact analytic expressions
for the polarizabilities as a function of $Z$ with a
relative accuracy of 10$^{-6}$ up to $Z = 100$ are determined by
fitting to the calculated polarizabilities.  The oscillator strengths
satisfy the sum rules $\sum_i f^{(\ell)}_{0i} = 0$ for all
multipoles from $\ell = 1$ to $\ell = 4$. The dispersion
coefficients for the long-range H-H and H-He$^+$ interactions are
given.

\end{abstract}

 \maketitle

\section{Introduction}

The present paper reports calculations of the polarizabilities of
the hydrogen atom and isoelectronic ions using the Dirac equation
to describe the underlying dynamics.  Such calculations are now
topical since some atomic polarizabilities can directly impact the
definitions of two fundamental quantities, the Kelvin and the
second~\cite{mitroy10a}. The new generation of optical frequency
standards have reached such precision that they are sensitive to
the black-body radiation of the apparatus itself \cite{gill05a}.
The resulting black-body radiation shift is largely determined by
the differences in polarizabilities of the two atomic states
involved in the clock transition.  Additionally, very high precision
measurements of the helium dielectric constant have been recently reported
\cite{schmidt07a}. In conjunction with high precision calculations
of the static dipole polarizability \cite{lach04a}, these
measurements can result in improved determinations of Boltzmann's
constant and thus the Kelvin \cite{schmidt07a,fellmuth11a}.

Another reason for doing such calculations is that they can be used to verify
the accuracy of computational methods and tests of fundamental
theory.  The polarizabilities of the hydrogenic ions are properties of
the ground state of a set of systems that are often used to
test the fundamental principles of physics.  It is rather surprising
that the first explicit calculations of the quadrupole polarizabilities
of the hydrogenic ions based on the Dirac equation have only just been reported
\cite{zhang12b}.

An important advance in the topic of the dipole and higher multipole
polarizabilities was an investigation based on the Pauli approximation
that gave expressions for the static multipole polarizabilities up to
$O(\alpha^2 Z^2)$ ~\cite{kaneko77a}.  This was a generalization of
an earlier work which gave the $O(\alpha^2 Z^2)$ expression for the
dipole polarizability \cite{bartlett69a}.  The work on dipole polarizabilities
was extended to $(\alpha Z)^4$ \cite{zon72a}, $(\alpha Z)^6$ \cite{yakhontov02a}
and to all orders in terms of a generalized hypergeometric function
\cite{szmytkowski04a}.  Apart from a very recent
calculation~\cite{zhang12b}, the expressions for the quadrupole and
higher-order polarizabilities have not had independent confirmation. There
have been a number of independent calculations of the dipole
polarizability and related sum rules.  Many of these investigations
have been computational in nature.  Drake and Goldman derived
expressions for some dipole oscillator strength sum rules as well as
performing some explicit calculations of the dipole polarizability
\cite{drake81a} by expanding the wave function as a linear
combination of exponential type functions.  Goldman~\cite{goldman89b} extended the
basis set approach to calculate the dipole polarizability of
hydrogenic ions from $Z=1$ to $Z=115$ using a Gauge-invariance
method. A fit to the calculated polarizabilities was used to create
an $(\alpha Z)^n$ expansion of the polarizability
including terms up to $(\alpha Z)^8$. There have been a number of
other computational investigations of the dipole polarizabilities of
hydrogenic ions based on Dirac equation
\cite{johnson88a,baye08a,thu94a,szmytkowski97a,bhatti03a,shabaev04a,beloy08a}.

The present calculations used the B-spline Galerkin method with Notre Dame
(ND) boundary conditions \cite{johnson88a}.  Other approaches to
the B-spline boundary conditions have been proposed
\cite{shabaev04a,beloy08a,fischer09a,grant09a,sun11a}.
There is at present no overwhelming reason for adopting more complicated
boundary conditions in preference to the ND boundary conditions.
The B-spline approach to atomic structure has a number of advantages
\cite{sapirstein96a,kang06a},
it does not lead to linear dependence, the basis can be made effectively
complete in a finite region of space, the details of the basis are easily
adjustable and results are numerically stable.  However, like all
basis set approaches, the  energy spectrum also has a sea of negative-energy
states (the Dirac sea) and it is also possible for spurious states to
appear in the positive energy spectrum~\cite{beloy08a}.  These issues
have been discussed extensively~\cite{johnson88a,sapirstein96a,shabaev04a,beloy08a}

The present B-spline calculations of the multipole polarizabilities
give numerical values that are more precise than any previous
calculation. Values of associated oscillator strength sum-rules are also
given.  The nuclear mass was set to be infinite and the point
nucleus model was adopted.  Values are reported for intermediate
sums including the entire set of states and also for a set of
calculations that omitted the negative energy states from the
Dirac Sea.  Analytic expressions for the
polarizabilities are constructed that are accurate to a relative
precision of $10^{-6}$ for $Z \le 100$.  The static multipole
polarizabilities for quadrupole, octupole and hexadecupole
transitions have been computed and found to be compatible with the
$O(\alpha^2 Z^2)$ expressions of Kaneko {\em et al} \cite{kaneko77a}.
The sum rules $\sum_i f^{(\ell)}_{0i} = 0$, provide a valuable
consistency check on the reliability of our calculations.
Finally, the dispersion
coefficients that describe the long-range interaction of the H-H and
H-He$^+$ dimers in their ground states are presented.  All results
are reported in atomic units and the value of fine structure
constant, $1/\alpha=c=137.035\ 999\ 074$~\cite{nist10} was used in all
calculations reported in this work unless specifically mentioned.

\section{Formulation}

\subsection{Dirac equation of single-electron atomic system}
The single-electron Dirac equation is
\begin{eqnarray}
H_{D}\psi(\textbf{r})=E\psi(\textbf{r})\,, \label {eq:t8}
\end{eqnarray}
$H_{D}$ is the Dirac Hamiltonian,
\begin{eqnarray}
H_{D}=c\bm{\alpha}\cdot\bm{p}+\beta mc^{2}+V(\textbf{r}) \,, \label
{eq:t9}
\end{eqnarray}
where $m$ is the electron mass, $c$ is the light velocity, $\bm{p}$
is the momentum operator, $\bm{\alpha}$ and $\beta$ are $4\times 4$
matrices of the Dirac operators~\cite{kaneko77a}.

The wavefunction for the hydrogen-like ion can be written
\begin{eqnarray}
\psi(\textbf{r})=\frac{1}{r} \left (
\begin{array}{r}
iP_{n\kappa}(r)\Omega_{\kappa m}(\hat{\textbf{r}})\\
Q_{n\kappa}(r)\Omega_{-\kappa m}(\hat{\textbf{r}})\\
\end{array}
\right ) \,, \label {eq:t7}
\end{eqnarray}
where $P_{n\kappa}(r)$ and $Q_{n\kappa}(r)$ present the larger and
small components of radial wavefunction, and $\Omega_{\kappa
m}(\hat{\textbf{r}})$ and $\Omega_{-\kappa m}(\hat{\textbf{r}})$ are
corresponding to the angular components. The angular quantum number
$\kappa$ are connected with $j$ and $\ell$,
\begin{equation}
\kappa=\ell(\ell+1)-j(j+1)-1/4 \,, \label{kappa}
\end{equation}
Substituting Eqs.~(\ref{eq:t9}) and (\ref{eq:t7})
into Eq.~(\ref{eq:t8}) and separating
the radial and angular components, gives the following coupled first-order
differential equations for radial components $P_{n\kappa}(r)$ and
$Q_{n\kappa}(r)$,
\begin{eqnarray}
[V(r)+mc^{2}]P_{n\kappa}(r)+c \left(\frac{d}{dr}-\frac{\kappa}{r} \right)Q_{n\kappa}(r)&\!=\!&EP_{n\kappa}(r) \,, \\
-c \left
(\frac{d}{dr}+\frac{\kappa}{r}\right)P_{n\kappa}(r)+[V(r)-mc^{2}]Q_{n\kappa}(r)&\!=\!&EQ_{n\kappa}(r)
\,, \label {eq:t10}
\end{eqnarray}
In this equation $V(r)$ is the interaction potential between the electron and
nucleus,
\begin{eqnarray}
V(r)=-\frac{Z}{r} \,,
\end{eqnarray}
with $Z$ being the number of nuclear charges.

In order to compare with non-relativistic calculations, we replace
the energy $E$ by $\varepsilon=E-mc^{2}$, and the radial Dirac
equation can be written as matrix style,
\begin{eqnarray}
\left (
\begin{array}{cc}
V(r)                              & c(\frac{d}{dr}-\frac{\kappa}{r})\\
-c(\frac{d}{dr}+\frac{\kappa}{r})  &-2mc^{2}+V(r)\\
\end{array}
\right )
\left (
\begin{array}{c}
P_{n\kappa}(r)\\
Q_{n\kappa}(r)\\
\end{array}
\right )
=\varepsilon
\left (
\begin{array}{c}
P_{n\kappa}(r)\\
Q_{n\kappa}(r)\\
\end{array}
\right ) \,. \label {eq:t11}
\end{eqnarray}

\subsection{B-spline Galerkin method }

The radial wavefunctions $P_{n\kappa}(r)$ and $Q_{n\kappa}(r)$ are
expanded in a $N$-dimensional basis of B-splines of order $k$,
\begin{eqnarray}
P(r)&=&\sum_{i=1}^{N}p_{i}B_{i}^{k}(r) \,,  \\
Q(r)&=&\sum_{i=1}^{N}q_{i}B_{i}^{k}(r) \,, \label {eq:t15}
\end{eqnarray}
where the subscripts $n, \kappa$ have been omitted from the
functions $P_{n\kappa}(r)$ and $Q_{n\kappa}(r)$ for notational
simplicity. The function $B_{i}^{k}(r)$ only take nonzero values for
the knot intervals $t_i\leq r \leq t_{i+k}$. The normalization
condition is
\begin{eqnarray}
\int_{0}^{\infty}[P^{2}(r)+Q^{2}(r)]dr=1 \,. \label {eq:t16}
\end{eqnarray}

The details of the B-splines and ND boundary conditions have been
discussed in detail elsewhere \cite{johnson88a,beloy08a}.
The large and small radial components are independently expanded
in a B-spline basis with the boundary conditions, $P(R)=Q(R)$
and $P(0)=0$, where $R$ is the radius of
confining cavity.

B-splines of $k=9$ order were used with the endpoints
of multiplicity $9$. An exponential knot distribution for
the B-splines is adopted, e.g.
\begin{eqnarray}
t_{i+k-1}=R\times\frac{\exp({\gamma(\frac{i-1}{N_1-1})})-1}{\exp({\gamma})-1}
\,, \label {eq:knots}
\end{eqnarray}
where $i=1, 2, \cdot\cdot\cdot, N_1$ and $N_1=N-k+2$ being the
maximal value of $i$. The exponential knot parameter $\gamma$
depends on the radius of confining cavity $R$,
\begin{eqnarray}
\gamma=G(Z)\times R \,.
\end{eqnarray}
The function $G(Z)$ for $Z \ge 2$ satisfies the recurrence relation
\begin{eqnarray}
G(Z)=G(Z-1)+\frac{0.055}{Z} \,,
\end{eqnarray}
where $G(1) = 0.055$ is an optimized value for the hydrogen atom. The
confining cavity radius $R$ (which is different for different $Z$)
was chosen to reproduce the exact ground-state energy \cite{bethe77a} of
the hydrogen-like ions to at least $20$ significant digits
\begin{eqnarray}
\varepsilon_n^{\rm Exact}=c^2\left[1+\frac{(\alpha
Z)^2}{[n-|\kappa|+\sqrt{\kappa^2-(\alpha Z)^2}]^2}\right]^{-1/2}\!-c^2
\,.\label{Eexact}
\end{eqnarray}
where $n$ is the main quantum number.

\subsection{Polarizabilities for the single-electron atoms}

In an weak external electric field, the static $2^{\ell}$-pole
polarizability for an atom is usually defined in
terms of a sum over all intermediate states including the continuum,
\begin{eqnarray}
\alpha_{\ell}=\sum_{i}\frac{f_{gi}^{(\ell)}}{(E_{i}-E_{g})^{2}}
\,.  \label {eq:alpha}
\end{eqnarray}
The initial state, $\psi_{g}(\textbf{r})$, with energy, $E_g$, is excluded from the
summation over $i$,
The $2^\ell$-pole oscillator strength $f_{gi}^{(\ell)}$ from ground
state $g$ to excited state $i$ is defined
\begin{eqnarray}
f_{gi}^{(\ell)}=\frac{2(E_{i}-E_{g})|\langle\psi_{g}(\textbf{r})\|r^{\ell}\textbf{C}^{(\ell)}(\hat{\textbf{r}})\|\psi_{i}(\textbf{r})\rangle|^{2}}{(2\ell+1)(2j_{g}+1)}
\,, \label {eq:t6}
\end{eqnarray}
where $j_{g}$ is the total angular momentum for the ground-state.
The wavefunction and energy of the excited states
are $\psi_{i}(\textbf{r})$ and
 $E_{i}$.
$\textbf{C}^{(\ell)}(\hat{\textbf{r}})$ is the $\ell$-order
spherical tensor.

Using Eq.~(\ref{eq:t7}), the radial and angular parts of matrix
element in the Eq.~(\ref{eq:t6}) are
\begin{eqnarray}
\langle \psi_{g}(r) | r^\ell |\psi_{i}(r)\rangle
=\int_{0}^{\infty}r^\ell[P_{g}(r)P_{i}(r)+Q_{g}(r)Q_{i}(r)]
d r \,, \label {eq:t12}
\end{eqnarray}
\begin{eqnarray}
\langle \Omega_{\kappa_g}(\hat{\textbf{r}})
\|\textbf{C}^{(\ell)}\|\Omega_{\kappa_i }(\hat{\textbf{r}})
\rangle&=&(-1)^{j_{g}+\frac{1}{2}}\sqrt{(2j_{g}+1)(2j_{i}+1)} \nonumber \\
&\times  &\left (
 \begin{array}{ccc}
j_g & j_i & \ell \\
-1/2 & 1/2 & 0 \\
\end{array}
\right ) \,. \label {eq:t13}
\end{eqnarray}
Polarizabilities that are computed including both the physical states and
negative energy states of the Dirac sea in Eq.~(\ref{eq:alpha}) are denoted by
$\alpha^{\pm}_{\ell}$.
Polarizabilities that are computed by omitting the negative energy states of
the Dirac Sea in Eq.~(\ref{eq:alpha}) are denoted by $\alpha^{+}_{\ell}$.
The states of the Dirac sea are energetically distinct from the physical states
The polarizabilities computed using the $O(\alpha^2 Z^2)$ expressions of
\cite{kaneko77a} are denoted as $\alpha^{\rm K}_{\ell}$.

The polarizabilities can be expanded as a series in powers of
$(\alpha Z)^2$.  The series is written
\begin{eqnarray}
\alpha_{\ell}^{\rm R}=\alpha_{\ell}^{\rm NR} \left[ 1+\sum_{i=1}^n
\lambda_{2i}(\alpha Z)^{2i} \right] \,, \label{powerseries}
\end{eqnarray}
where the non-relativistic multipole polarizabilities,
$\alpha_{\ell}^{\rm NR}$,
for the ground-state hydrogen-like ions,
which have the exact values~\cite{dalgarno55a}
\begin{equation}
\alpha_{\ell}^{\rm NR} =
\frac{(2\ell+2)!(\ell+2)}{\ell(\ell+1)2^{2\ell+1}Z^{2\ell+2}} \,.
\end{equation}

\subsection{Oscillator strength sum rules}

There are a number of oscillator strength sum rules besides those
which define the multipole polarizabilities.  We make the definition
\begin{equation}
S_\ell(n) = \sum_i f_{gi}^{(\ell)}(E_i-E_g)^n \,. \label{fsum}
\end{equation}
 The expression with $\ell = 1$ and $n=-2$ is the dipole polarizability,
The case when $\ell = 1$ and $n = 0$ is called the
Thomas-Reiche-Kuhn (TRK) sum rule. In the non-relativistic calculation, $S_1(0)$
should be equal to the number of the electrons.
The case with $\ell = 1$
and $n = -3$ is related to the non-adiabatic dipole polarizability
\cite{dalgarno68a}. One finds that $S_1(-3) = 43/(4Z^6)$ for
non-relativistic hydrogenic atoms.
The $S_1(-1)$ coefficient \cite{lamm77b} is related to the
long-range atom wall dispersion coefficient \cite{mitroy03f}.
One finds that $S_1(-1) = 2/Z^2$ for non-relativistic hydrogenic
atoms.   The relativistic
sum rules are useful in testing the completeness of basis sets for
variational representations of the Dirac spectrum~\cite{goldman82a}
and set a foundation for testing other methods.

As with the polarizabilities, the sum-rules can be evaluated by summing
over all states, or just the positive energy states.
Sum rules that are computed including both positive and negative energy
states in Eq.~(\ref{fsum}) are denoted by $S^{\pm}_{\ell}(n)$.
Sum rules that omit the states from the negative energy sea from
the sum are denoted by $S^{+}_{\ell}(n)$.

\section{Results and Discussions}

\subsection{Polarizabilities and sum rules for hydrogen}

\begin{figure}[tbh]
\centering{
\includegraphics[width=8.4cm]{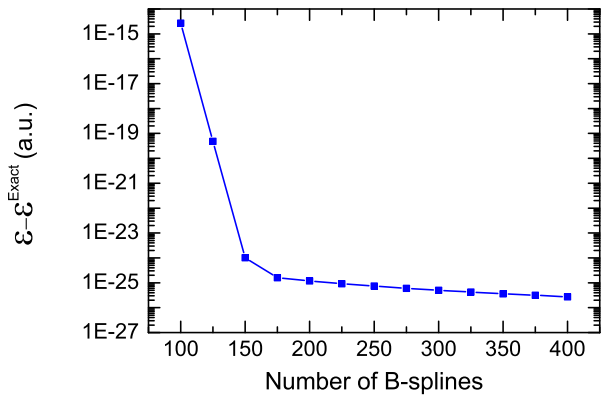}
} \caption{ \label{Fig1} (color online). The convergence of
ground-state energy (a.u.) relative to the exact Dirac equation
energy of the hydrogen ground state.  The number of B-spline basis
functions is $N$, while the radius of confining cavity is $R=400$ a.u. }
\end{figure}

\begingroup
\squeezetable
\begin{table*}
\caption{\label{polarizabilitiesH} The convergence of the static
multipole polarizabilities $\alpha^{\pm}_{\ell}$ (units of a.u.)
for the hydrogen atom ground state as the dimension,
$N$, of the B-spline basis set was increased.  The radius
of the confining cavity is $R=400$
 a.u.}
\begin{ruledtabular}
\begin{tabular}{lllll} \multicolumn{1}{c}{N}&
\multicolumn{1}{c}{$\alpha^{\pm}_1$} & \multicolumn{1}{c}{$\alpha^{\pm}_2$} &
\multicolumn{1}{c}{$\alpha^{\pm}_3$} & \multicolumn{1}{c}{$\alpha^{\pm}_4$} \bigstrut[b] \\
\hline
 100   &4.499\ 751\ 495\ 18               &14.998\ 829\ 821             &131.2379                & \bigstrut[t] \\
 150   &4.499\ 751\ 495\ 177\ 64          &14.998\ 829\ 822\ 856\ 41        &131.237\ 821\ 447\ 83         &2126.028\ 674\ 4992\\
 200   &4.499\ 751\ 495\ 177\ 639\ 27       &14.998\ 829\ 822\ 856\ 441\ 76     &131.237\ 821\ 447\ 844\ 63      &2126.028\ 674\ 499\ 1281\\
 250   &4.499\ 751\ 495\ 177\ 639\ 267\ 48    &14.998\ 829\ 822\ 856\ 441\ 70     &131.237\ 821\ 447\ 844\ 661     &2126.028\ 674\ 499\ 128\ 81\\
 300   &4.499\ 751\ 495\ 177\ 639\ 267\ 398   &14.998\ 829\ 822\ 856\ 441\ 699\ 67  &131.237\ 821\ 447\ 844\ 662\ 144  &2126.028\ 674\ 499\ 128\ 831\ 0\\
 350   &4.499\ 751\ 495\ 177\ 639\ 267\ 396\ 1  &14.998\ 829\ 822\ 856\ 441\ 699\ 61  &131.237\ 821\ 447\ 844\ 662\ 150\ 7 &2126.028\ 674\ 499\ 128\ 831\ 4\\
 400   &4.499\ 751\ 495\ 177\ 639\ 267\ 396\ 02 &14.998\ 829\ 822\ 856\ 441\ 699\ 608 &131.237\ 821\ 447\ 844\ 662\ 151\ 0 &2126.028\ 674\ 499\ 128\ 831\ 46\\
\end{tabular}
\end{ruledtabular}
\end{table*}
\endgroup

The difference of the B-spline ground-state energy from the exact
energy given by Eq.~(\ref{Eexact}) (this is
$-0.500\ 006\ 656\ 596\ 553\ 596\ 900\ 786\ 4298$ a.u.) as a function of the
dimension of the B-spline basis is plotted in Fig.\ref{Fig1}.  This
calculation was performed with a confinement radius of $R = 400$ a.u..
This ensures that none of the atomic sum rules reported
in this paper are influenced by the size of the confinement radius.
The energy was converged to 25 significant digits for a basis with
$N = 400$.

Table~\ref{polarizabilitiesH} shows the convergence of the static
multipole polarizabilities, $\alpha_\ell^{\pm}$, for the H$(1s)$
state as the dimension of the B-spline basis was increased
from $N = 100$ to $N = 400$. The radius of confining cavity is
$R=400$ a.u. The static dipole polarizability $\alpha_1^{\pm}$, is
computed to a precision of $22$ digits. The higher-order
polarizabilities $\alpha_2^{\pm}$, $\alpha_3^{\pm}$, and
$\alpha_4^{\pm}$ have not achieved the same degree of precision, but
are still computed to a precision of $21$, $20$, and $20$ effective
figures respectively. The present $\alpha_1^{\pm}$ =
$4.499\ 751\ 495\ 177\ 639\ 267\ 396\ 02$ a.u. is $4\times 10^{-11}$ a.u. larger
than the result $4.499\ 751\ 495\ 142\ 92$ a.u. of
Goldman~\cite{goldman89b}.  This is due to the different fine structure
constant used.  When the fine-structure constant $\alpha$, is set to
the value used by Goldman, namely $1/\alpha=137.035\ 999\ 074$, the
B-spline polarizability changed to $\alpha_1^{\pm}$ =
$4.499\ 751\ 495\ 142\ 916$ a.u. This is in perfect agreement with that of
Goldman.   All hydrogen atom sum-rules reported from now on use the
$N = 400$, $R = 400$ a.u. B-spline basis.

\begingroup
\squeezetable
\begin{table*}
\caption{\label{polarH1} The comparison of dipole sum rules, $S^{\pm}_1(n)$
and $S^{+}_1(n)$, for the H($1s$) ground state. The exact expressions of
sum rule are also presented in the fourth-column with
$\gamma_1=\sqrt{\kappa^2-\alpha^2Z^2}$~\cite{drake81a}. The ratio
$\Delta S_1(n)/S_1^{\rm exact}(n)=[S_1^{\rm exact}(n)-S^{\pm}_1(n)]/S_1^{\rm exact}(n)$.
The non-relativistic values are in the column $S_1^{\rm NR}(n)$.
The value of $S^{\pm}_1(0)$ is not stable and gets smaller
as the B-spline basis dimension is increased.  The notation $a[b]$ means $a \times
10^b$.}
\begin{ruledtabular}
\begin{tabular}{lcccccccc}
\multicolumn{1}{l}{Sum rule}& \multicolumn{1}{c}{$S^{+}_1(n)$}  &
\multicolumn{1}{c}{$S^{\pm}_1$(n)} &
\multicolumn{1}{c}{$S^{\rm NR}_1(n)$} &
\multicolumn{1}{c}{$S^{\rm exact}_1$(n)} &
\multicolumn{1}{c}{$\displaystyle{\frac{\Delta S_1(n)}{S_1^{\rm exact}(n)}}$}
\\
\hline
$S_1(-3)$               &10.749\ 260\ 777\ 454\ 106\ 9          &10.749\ 260\ 777\ 454\ 125\ 8       &  10.75 &                     & \bigstrut[t] \\
$S_1(-2)$               &4.499\ 751\ 495\ 886\ 496\ 76          &4.499\ 751\ 495\ 177\ 639\ 27       &   4.50  &  \\
$S_1(-1)$               &1.999\ 911\ 249\ 278\ 034\ 15          &1.999\ 937\ 873\ 065\ 244\ 31       &  2.0   & $\displaystyle{\frac{(\gamma_1+1)(2\gamma_1+1)}{3Z^2}}$              & 7[$-$19]     \\
$S_1(0)$                &0.999\ 955\ 631\ 350\ 807\ 45          &1[$-$29]  & 1.0 &  0           &$-$1[$-$29] \\
$S_1(1)$                &0.666\ 563\ 210\ 276\ 996\ 94          &3.755\ 773\ 008\ 441\ 865\ 7y[4]   &  $\displaystyle{\frac{2}{3}}$ & $\displaystyle{2/\alpha^2}$                                                                               &2[$-$18] \\
$S_1(2)$                &1.298\ 802\ 722\ 313          &$-$1.410\ 595\ 609\ 170\ 78[9] & $\displaystyle{\frac{4}{3}}$   &$\displaystyle{-\frac{4}{3\alpha^4}\left(\gamma_1+\frac{2}{\gamma_1}\right)}$                                         &7[$-$16] \\
$S_1(3)$                &---                          &5.298\ 0179\ 899\ 7[13]  & ---   & $\displaystyle{\frac{8}{3\alpha^6}\left[\frac{2(\gamma_1^2-1)(\gamma_1-2)}{\gamma_1(2\gamma_1-1)}+3\right]}$ &$-$2[$-$12] \\
\end{tabular}
\end{ruledtabular}
\end{table*}
\endgroup

Exact expressions exist for a number of the dipole sum rules given by
Eq.~(\ref{fsum}).  For example, the expressions for the exact non-relativistic
electric-dipole sum rules $S_1(n)$ have been derived for
$n=-5,-4,\cdots,2$~\cite{bell67a,lamm77b}. The non-relativistic dipole
sum-rule diverges for $n \ge 3$.
Expressions for some dipole sum rules for the Dirac hydrogen atom have been
given by Drake and Goldman~\cite{drake81a}.   The Dirac equation sum rules
were derived by using closure to sum over the complete set of positive and
negative energy
states and the expressions are given in Table \ref{polarH1}.  The Dirac
equation sum-rule for $S_1(3)$ is convergent due to cancellations between
the terms with positive and negative energies.

Table~\ref{polarH1} compares the dipole sum rules of the H($1s$) with
and without the contributions of the states in the negative energy sea.
All the digits listed are converged with respect to further enlargement
of the B-spline basis.  The sum-rules, $S^{\pm}_1(0)$, $S^{\pm}_1(1)$ and
$S^{\pm}_1(2)$, agree with the exact expressions to
better than 15 digits. Agreement is not so good for $S_1(3)$ but
in this case the sum is more sensitive to terms that occur at larger
positive and negative energies.  There was no evidence of convergence
for $S_1(3)$ when the states of the negative energy sea were omitted
from intermediate sum.  This is consistent with the non-relativistic result
of Lamm and Szabo~\cite{lamm77a}.

\begingroup
\squeezetable
\begin{table*}
\caption{\label{polarH2} The comparison of the H$(1s)$ static multipole
polarizabilities and sum rules with and without the negative energy
states. Values for $S^{\pm}_{\ell}(0)$ are not numerically stable and
tend to decrease as the basis is enlarged.
The notation $a[b]$ means $a \times 10^b$.}
\begin{ruledtabular}
\begin{tabular}{llll}
\multicolumn{1}{l}{Sum rule}&
\multicolumn{1}{c}{$S^{+}_{\ell}$ }& \multicolumn{1}{c}{$S^{\pm}_{\ell}$ }
& \multicolumn{1}{c}{Non-relativistic} \\
\hline
$S_2(-3)$                &26.747\ 582\ 450\ 922\ 508\ 1          &26.747\ 582\ 450\ 922\ 621\ 3 &   26.750  \bigstrut[t] \\
$S_2(-2)$                &14.998\ 829\ 827\ 109\ 609\ 3          &14.998\ 829\ 822\ 856\ 441\ 7 &   15.0 \\
$S_2(-1)$                &8.999\ 384\ 961\ 848\ 033\ 62          &8.999\ 544\ 703\ 293\ 683\ 54 &    9.0  \\
$S_2(0)$                 &5.999\ 605\ 961\ 023\ 935\ 20          &$~-$$1[-28]$ & 6.0    \\
$S_2(1)$                 &4.799\ 574\ 122\ 244\ 155\ 50          &2.253\ 393\ 804\ 665\ 723\ 66[5] &  4.80   \\
$S_2(2)$                 &5.598\ 084\ 401\ 298\ 530\ 18          &$-$8.463\ 168\ 026\ 239\ 51[9] & 5.60  \\
$S_2(3)$                 &20.043\ 653\ 259\ 626\               &3.178\ 616\ 122\ 279\ 928[14]  & 20.80  \\
\hline
$S_3(-3)$                &204.041\ 400\ 069\ 326\ 002          &204.041\ 400\ 069\ 327\ 276   & 204.06250  \bigstrut[t] \\
$S_3(-2)$                &131.237\ 821\ 495\ 692\ 427          &131.237\ 821\ 447\ 844\ 662   &  131.250 \\
$S_3(-1)$                &89.992\ 366\ 277\ 948\ 754\ 1          &89.994\ 163\ 347\ 335\ 786\ 6   &  90.0  \\
$S_3(0)$                 &67.494\ 445\ 945\ 723\ 638\ 3          &$~6[-24]$ & 67.50 \\
$S_3(1)$                 &57.851\ 751\ 717\ 231\ 297\ 2          &2.535\ 018\ 531\ 367\ 228\ 21[6] &  $\frac{405}{7} = 57.8571428$ \\
$S_3(2)$                 &61.704\ 978\ 988\ 083\ 113\ 1          &$-$9.520\ 821\ 387\ 020\ 48[10] & $\frac{432}{7} = 61.7142857$  \\
$S_3(3)$                 &100.225\ 824\ 655\ 056\ 308          &3.575\ 836\ 136\ 347\ 471[15]  & $\frac{702}{7} = 100.2857142$    \\
\hline
$S_4(-3)$                &3043.342\ 638\ 220\ 471\ 07          &3043.342\ 638\ 220\ 494\ 85 & 3043.687\ 50 \bigstrut[t] \\
$S_4(-2)$                &2126.028\ 675\ 392\ 279\ 56          &2126.028\ 674\ 499\ 128\ 83 & 2126.25 \\
$S_4(-1)$                &1574.846\ 608\ 527\ 950\ 22          &1574.880\ 153\ 465\ 178\ 52 & 1575.0  \\
$S_4(0)$                 &1259.880\ 083\ 503\ 994\ 17          &$~9[-19$] &  1260.0  \\
$S_4(1)$                 &1119.885\ 837\ 666\ 203\ 30          &4.731\ 967\ 094\ 641\ 248\ 23[7] &  1120.0    \\
$S_4(2)$                 &1159.848\ 826\ 903\ 744\ 95          &$-$1.777\ 189\ 196\ 560\ 64[12] & 1160.0  \\
$S_4(3)$                 &1530.311\ 794\ 804\ 461              &6.674\ 763\ 648\ 090\ 144[16]  &  $\frac{4591}{3} =  1530.66666666$ \\
\end{tabular}
\end{ruledtabular}
\end{table*}
\endgroup

The value $S^{+}_1(0)$, which omits the states from the Dirac sea,
is close to the non-relativistic value of nuclear charge
$Z=1$.  Upon making the substitution $\langle p^2 \rangle = Z^2$
in existing expressions \cite{levinger57a,cohen05b}, we obtain the
result
\begin{equation}
S^{+ {\rm Levinger}}_1(0) = 1 - \frac{5 \alpha^2 Z^2}{6} + \ldots \ ,
\end{equation}
Evaluating this expression for $Z=1$ gives, $S_1(0) = 0.999\ 955\ 6238$,
which is only $8 \times 10^{-9}$ different from the B-spline evaluation.
The degree of difference between $S^{\rm NR}_1(n)$ and $S^{\rm +}_1(n)$
gets larger as $n$ increases. The difference is $3_0\%$ for $S_1(2)$.

The contribution that the negative-energy states make to the dipole sum
rules depends on $n$.  The negative-energy states of the Dirac sea contribute
less than $2\times 10^{-5}$ to $S^{\pm}_1(-1)$, $2 \times 10^{-10}$ to
$S^{\pm}_1(-2)$, and $2\times 10^{-15}$ to $S^{\pm}_1(-3)$.  This is not
surprising.  The negative energy states are located at energies
of order $-$$2c^2$.   So the contributions of the negative energy
states decrease as $n$ in Eq.~(\ref{fsum}) becomes
increasingly negative. Conversely, the differences between
the $S^{\pm}_1(n)$ and the $S^{+}_1(n)$ sum rules can be
expected to increase as $n$ increases.   Table \ref{polarH1}
shows that this indeed does happen.  The difference between
$S^{\pm}_1(2)$ and $S^{+}_1(2)$ is nine orders of
magnitude.

Table~\ref{polarH2} gives the sum rules for the higher-order
multipoles for the hydrogen-atom ground state. The $S_2(-2)$,
$S_3(-2)$, and $S_4(-2)$ are the multipole polarizabilities
$\alpha_2$, $\alpha_3$, and $\alpha_4$
respectively. The sum-rules, $S^{+}_{\ell}(n)$, omitting the states
from the Dirac sea
are within 0.1$\%$ of the non-relativistic values with the exception
of $S_2(3)$. This is also true
for the sum-rules, $S^{\pm}_{\ell}(n)$, with $n < 0$ that also include the
Dirac sea.

The most striking results from Table \ref{polarH2} are the $S^{\pm}_{\ell}(0)$
sum-rules which do not exceed $10^{-18}$.  Levinger {\em et al} \cite{levinger57a}
have pointed out that the Dirac Hamiltonian involves terms linear in the
particle momentum ${\bf p}$ and that as a consequence the Bethe sum rule
for $\exp(i {\bf q} \cdot {\bf r})$ should be identically zero.  The expansion
of $\exp(i {\bf q} \cdot {\bf r})$ implicitly involves dipole, quadrupole
and octupole matrix elements.  Therefore, it is expected that
$S_{\ell}(0) = 0$ for all $\ell$.

The contributions of the negative-energy Dirac Sea to the $S^{\pm}_{\ell}(n)$
sum-rules are actually greater than the contributions from the physical states
for $n \ge 1$.  They exceed the contribution from the physical states by amounts
from $4$ to $14$ orders of magnitude.

\subsection{Polarizabilities for the hydrogen isoelectronic series}

\begingroup
\squeezetable
\begin{table*}
\caption{\label{hydrogenlike1} The static dipole polarizabilities
for the ground-state of selected hydrogen-like
ions. The present values are listed in the third- and fourth-columns
for two sets of $c=1/\alpha$. All the tabulated digits of the
present work are insensitive to further enlargement of the basis.
The notation $a[b]$ means $a \times 10^b$.}
\begin{ruledtabular}
\begin{tabular}{lcccc}
 \multicolumn{1}{l}{$Z$}&   $\alpha_1^+$     & $\alpha_1^{\pm}$   & $\alpha_1^{\pm}$   & $\alpha_1^{\pm}$\\
  &  \multicolumn{1}{c}{$c=137.035\ 999\ 074$}  &   \multicolumn{1}{c}{$c=137.035\ 999\ 074$}& \multicolumn{1}{c}{$c=137.035\ 989\ 5$}& \multicolumn{1}{c}{Goldman~\cite{goldman89b}}\\ \hline
1    & 4.499\ 751\ 495\ 886\ 496\ 765\ 8     & 4.499\ 751\ 495\ 177\ 639\ 267\ 4     & 4.499\ 751\ 495\ 142\ 915\ 967\ 2    & 4.499\ 751\ 495\ 142\ 92 \bigstrut[t] \\
2    & 0.281\ 187\ 875\ 627\ 153\ 384\ 5     & 0.281\ 187\ 874\ 918\ 503\ 235\ 4     & 0.281\ 187\ 874\ 909\ 822\ 724\ 5   & 0.281\ 187\ 874\ 909\ 82  \\
5    & 7.190\ 061\ 953\ 255\ 011\ 860[$-$3]  & 7.190\ 061\ 460\ 476\ 174\ 63[$-$3]  & 7.190\ 061\ 244\ 659\ 087\ 6[$-$3]   & 7.190\ 061\ 244\ 659\ 0[$-$3] \\
10   & 4.475\ 171\ 382\ 242\ 160\ 041[$-$4]  & 4.475\ 164\ 360\ 625\ 272\ 209[$-$4]  & 4.475\ 164\ 357\ 157\ 090\ 8[$-$4]   & 4.475\ 164\ 357\ 157[$-$4] \\
15   & 8.778\ 661\ 031\ 860\ 895\ 31[$-$5]   & 8.778\ 591\ 625\ 838\ 392\ 08[$-$5]   & 8.778\ 591\ 610\ 447\ 560\ 3[$-$5]   & 8.778\ 591\ 610\ 447[$-$5] \\
20   & 2.750\ 591\ 823\ 590\ 310\ 61[$-$5]   & 2.750\ 523\ 499\ 062\ 579\ 08[$-$5]   & 2.750\ 523\ 490\ 423\ 618\ 6[$-$5]   & 2.750\ 523\ 490\ 424[$-$5] \\
25   & 1.112\ 456\ 189\ 324\ 034\ 04[$-$5]   & 1.112\ 389\ 181\ 457\ 920\ 41[$-$5]   & 1.112\ 389\ 175\ 944\ 142\ 1[$-$5]   & 1.112\ 389\ 175\ 944[$-$5] \\
30   & 5.281\ 595\ 642\ 877\ 009\ 9[$-$6]    & 5.280\ 940\ 730\ 404\ 758\ 7[$-$6]    & 5.280\ 940\ 692\ 243\ 592\ 6[$-$6]   & 5.280\ 940\ 692\ 243[$-$6] \\
35   & 2.798\ 031\ 223\ 308\ 353\ 9[$-$6]    & 2.797\ 393\ 149\ 766\ 563\ 6[$-$6]    & 2.797\ 393\ 121\ 842\ 089\ 7[$-$6]   & 2.797\ 393\ 121\ 842[$-$6] \\
40   & 1.604\ 622\ 695\ 629\ 832\ 0[$-$6]    & 1.604\ 002\ 839\ 548\ 263\ 7[$-$6]    & 1.604\ 002\ 818\ 268\ 128\ 9[$-$6]   & 1.604\ 002\ 818\ 268[$-$6]  \\
45   & 9.767\ 839\ 136\ 814\ 269[$-$7]     & 9.761\ 833\ 945\ 433\ 110[$-$7]     & 9.761\ 833\ 778\ 188\ 453\ 4[$-$7]   & 9.761\ 833\ 778\ 187[$-$7] \\
50   & 6.226\ 889\ 347\ 856\ 944[$-$7]     & 6.221\ 086\ 480\ 106\ 640[$-$7]     & 6.221\ 086\ 345\ 451\ 685\ 9[$-$7]   & 6.221\ 086\ 345\ 451[$-$7]  \\
55   & 4.116\ 918\ 654\ 470\ 464[$-$7]     & 4.111\ 325\ 157\ 474\ 914[$-$7]     & 4.111\ 325\ 046\ 935\ 820\ 5[$-$7]   & 4.111\ 325\ 046\ 936[$-$7] \\
60   & 2.802\ 469\ 149\ 798\ 750[$-$7]     & 2.797\ 090\ 474\ 417\ 353[$-$7]     & 2.797\ 090\ 382\ 223\ 343\ 9[$-$7]   & 2.797\ 090\ 382\ 224[$-$7]  \\
65   & 1.953\ 091\ 120\ 155\ 380[$-$7]     & 1.947\ 931\ 407\ 519\ 126[$-$7]     & 1.947\ 931\ 329\ 604\ 639\ 5[$-$7]   & 1.947\ 931\ 329\ 604[$-$7] \\
70   & 1.387\ 222\ 340\ 637\ 801[$-$7]     & 1.382\ 284\ 686\ 111\ 543[$-$7]     & 1.382\ 284\ 619\ 529\ 769\ 8[$-$7]   & 1.382\ 284\ 619\ 530[$-$7]  \\
75   & 1.000\ 397\ 933\ 028\ 34[$-$7]      & 9.956\ 846\ 315\ 732\ 27[$-$8]      & 9.956\ 845\ 741\ 359\ 051\ 6[$-$8]   & 9.956\ 845\ 741\ 359[$-$8]  \\
80   & 7.301\ 102\ 574\ 925\ 93[$-$8]      & 7.256\ 230\ 363\ 582\ 21[$-$8]      & 7.256\ 229\ 864\ 059\ 587\ 8[$-$8]   & 7.256\ 229\ 864\ 060[$-$8] \\
85   & 5.376\ 751\ 290\ 435\ 41[$-$8]      & 5.334\ 153\ 759\ 283\ 73[$-$8]      & 5.334\ 153\ 321\ 793\ 719\ 5[$-$8]   & 5.334\ 153\ 321\ 795[$-$8]  \\
90   & 3.984\ 403\ 901\ 650\ 36[$-$8]      & 3.944\ 093\ 881\ 570\ 48[$-$8]      & 3.944\ 093\ 496\ 045\ 404\ 3[$-$8]   & 3.944\ 093\ 496\ 045[$-$8] \\
95   & 2.962\ 871\ 397\ 452\ 10[$-$8]      & 2.924\ 863\ 256\ 366\ 13[$-$8]      & 2.924\ 862\ 914\ 773\ 842\ 2[$-$8]   & 2.924\ 862\ 914\ 774[$-$8] \\
100  & 2.204\ 334\ 865\ 912\ 88[$-$8]      & 2.168\ 647\ 587\ 493\ 68[$-$8]      & 2.168\ 647\ 283\ 324\ 507\ 9[$-$8]   & 2.168\ 647\ 283\ 325[$-$8]\\
\end{tabular}
\end{ruledtabular}
\end{table*}
\endgroup

Table~\ref{hydrogenlike1} presents the static dipole polarizabilities
for a number for hydrogen-like ions in their ground-state.
All the digits listed in this table are converged with respect to further
enlargement of the B-spline basis.  In order to facilitate
comparison of the present polarizabilities with those of
Goldman~\cite{goldman89b}, we repeated the calculations but used the
the same speed of light $c=137.035\ 989\ 5$ (in atomic units) as Goldman.
The agreement with the polarizabilities of
Goldman could hardly have been better.  At $Z = 10$
we got $\alpha_1^{\pm} = 4.475\ 164\ 357\ 157\ 0908\times 10^{-4}$ a.u.,
in agreement with all published digits of Goldman.  The same is true for the
polarizability at $Z = 90$, namely
$3.944\ 093\ 496\ 045\ 4043 \times 10^{-8}$ a.u.
This level of agreement was
achieved for all values of $Z$ from 1 to 100.  The only
disagreements amounted to $\pm 1$ in the last significant digit
reported by Goldman~\cite{goldman89b}.

\begin{figure}[tbh]
\centering{
\includegraphics[width=8.4cm]{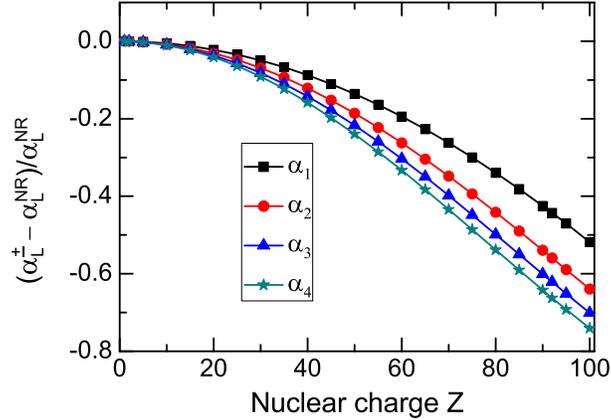}
} \caption{ \label{Fig2} (color online). The impact of relativistic
effects on the multipole polarizabilities for the hydrogen
isolectronic series.  The ratio $(\alpha_{\ell}^{\pm} - \alpha_{\ell}^{\rm
NR})/\alpha_{\ell}^{NR}$ is plotted. }
\end{figure}

The higher-order polarizabilities, $\alpha^{\pm}_{\ell}$ of
the ground-states of some selected hydrogen-like ions are presented in
Table~\ref{hydrogenlike2}. All the reported digits are insensitive to
further enlargement in the B-spline basis.
Fig.~\ref{Fig2} shows the influence of relativistic effect on multipole
polarizabilities. The relativistic effect becomes larger as the nuclear
charge, $Z$, is increased.
The relative size of the relativistic effect is smallest for the dipole
polarizability and largest for $\alpha_4$.

\begin{figure}[tbh]
\centering{
\includegraphics[width=8.4cm]{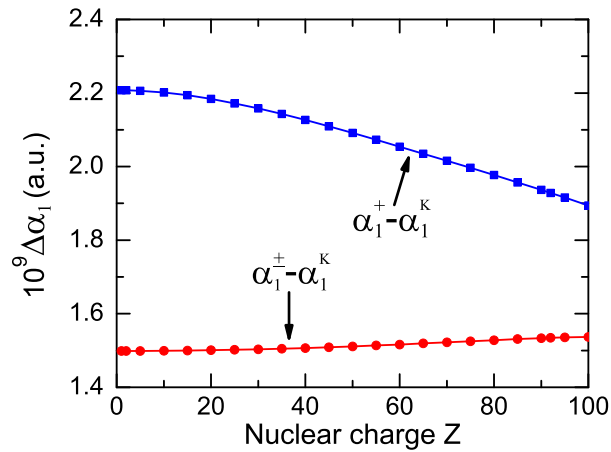}
} \caption{ \label{Fig3} (color online). Plot of $10^{9} \Delta
\alpha_1/Z^2$ as a function of nuclear charge, $Z$. }
\end{figure}

The difference of the $\alpha_1^{\pm}$ and $\alpha_1^{+}$
polarizabilities from the Kaneko polarizabilities are illustrated in
Fig.~\ref{Fig3}.  We define $\Delta \alpha_1^{\pm} = (\alpha_1^{\pm}
- \alpha_1^{\rm K})$ with a similar relation used to define
$\Delta\alpha_1^{+} $. Fig.~\ref{Fig3} plots $10^9 \Delta \alpha_1$
as a function of $Z$. These are seen to go to a constant value as $Z
\to 0$. From Eq.~(\ref{powerseries}) we deduce
\begin{equation}
\Delta \alpha_1^{\pm} = \frac{9}{2Z^4} \left[ \left(
\lambda_2+\frac{28}{27} \right) (\alpha Z)^2 +  O(\alpha^4 Z^4)
\right ]  .
\end{equation}
This expression can only go to a constant in the $Z \to 0$ limit
when $\lambda_2 = -\frac{28}{27}$. Fig.~\ref{Fig3} demonstrates that
$\alpha_1^{\pm}$, $\alpha_1^{+}$, and $\alpha_1^{\rm K}$ are equal
to order $O(\alpha^2 Z^2)$. The different $Z \to 0$ asymptotes for
$\Delta \alpha_1^{\pm} $ and $\Delta \alpha_1^{+} $ indicate that
the $O(\alpha^4 Z^4)$ terms are different for $\alpha_1^{\pm}$ and
$\alpha_1^{+}$.

\begin{figure}[tbh]
\centering{
\includegraphics[width=8.4cm]{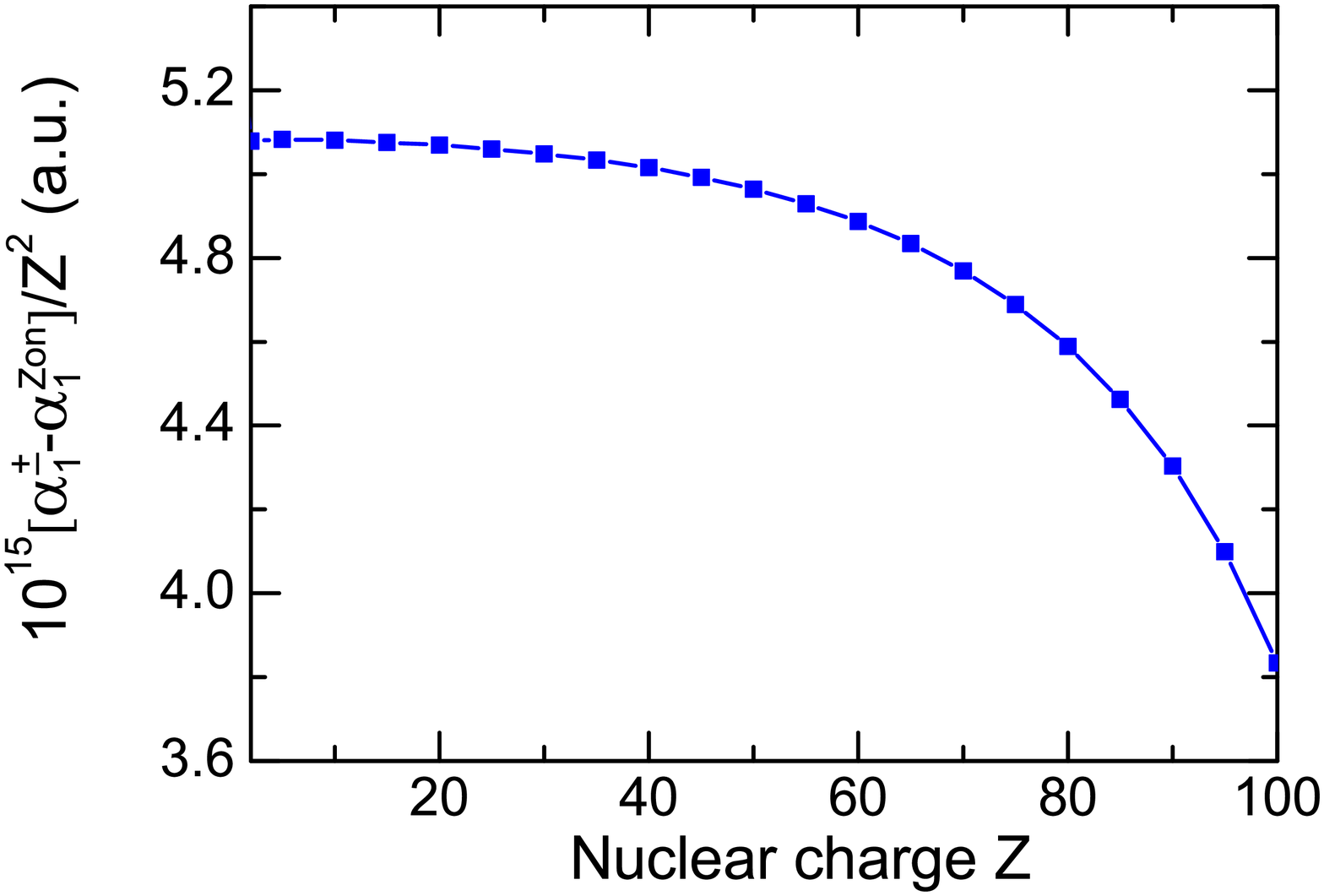}
} \caption{ \label{Fig4} (color online). Plot of
$10^{15} (\alpha_1^{\pm} - \alpha_1^{\rm Zon})/Z^2$
as a function of nuclear charge, $Z$. }
\end{figure}

Expressions for $\alpha^{\pm}_1$ giving terms to $O(\alpha^4Z^4)$
\cite{zon72a}, $O(\alpha^6Z^6)$ \cite{yakhontov02a} and to all
orders \cite{szmytkowski04a} have been
derived.  The $O(\alpha^4Z^4)$ expression of Zon \cite{zon72a} is
\begin{equation}
\alpha_1^{\rm Zon} = \frac{9}{2Z^4} \left[ 1 -
\frac{28}{27} (\alpha Z)^2 +  \frac{31+2\pi^2}{432} (\alpha^4 Z^4)
+ \ldots \right ]  .
\end{equation}
Figure \ref{Fig4} plots $10^{15}(\alpha_1^{\pm} - \alpha_1^{\rm Zon})/Z^2$
as a function of $Z$.  The $Z \to 0$ limit of this difference
demonstrates that the present calculations are in agreement with
the analytic expression to $O(\alpha^4Z^4)$.  This provides
a mutual validation of the B-spline calculations and the analytic
expressions.

\begin{table*}
\caption{\label{hydrogenlike2} Relativistic multipole
polarizabilities (in a.u.) for the ground states of the
hydrogen isoelectronic series.  All the figures listed are accurate.
The notation $a[b]$ means $a \times 10^b$.}
\begin{ruledtabular}
\begin{tabular}{cccc} \multicolumn{1}{c}{$Z$}& \multicolumn{1}{c}{$\alpha^{\pm}_2$}&
\multicolumn{1}{c}{$\alpha^{\pm}_3$}&
\multicolumn{1}{c}{$\alpha^{\pm}_4$}\\
\hline
1        &14.998\ 829\ 822\ 856\ 441\ 699        &131.237\ 821\ 447\ 844\ 662        &2126.028\ 674\ 499\ 128\ 83 \bigstrut[t] \\
2        &0.234\ 301\ 867\ 935\ 791\ 210\ 0        &0.512\ 505\ 037\ 523\ 770\ 47        &2.075\ 551\ 5460\ 612\ 051\ 9\\
5        &9.581\ 285\ 372\ 324\ 0453\ 92[$-$4]     &3.352\ 210\ 608\ 787\ 016\ 2[$-$4]     &2.171\ 618\ 426\ 945\ 541\ 1[$-$4]\\
10       &1.488\ 319\ 383\ 913\ 411\ 04[$-$5]      &1.300\ 352\ 899\ 787\ 624[$-$6]      &2.104\ 187\ 645\ 750\ 314[$-$7]\\
15       &1.293\ 852\ 351\ 688\ 892\ 4[$-$6]       &5.014\ 877\ 480\ 967\ 07[$-$8]       &3.601\ 503\ 954\ 501\ 5[$-$9]\\
20       &2.271\ 146\ 583\ 050\ 793[$-$7]        &4.938\ 640\ 072\ 269\ 2[$-$9]        &1.991\ 062\ 443\ 017[$-$10]\\
25       &5.847\ 845\ 585\ 737\ 33[$-$8]         &8.110\ 859\ 162\ 392[$-$10]        &2.087\ 370\ 771\ 99[$-$11]\\
30       &1.915\ 515\ 761\ 865\ 58[$-$8]         &1.837\ 296\ 630\ 650[$-$10]        &3.273\ 123\ 521\ 7[$-$12]\\
35       &7.397\ 473\ 245\ 589\ 1[$-$9]          &5.186\ 973\ 978\ 69[$-$11]         &6.763\ 105\ 560[$-$13]\\
40       &3.218\ 326\ 876\ 369\ 0[$-$9]          &1.717\ 671\ 116\ 72[$-$11]         &1.707\ 067\ 337[$-$13]\\
45       &1.531\ 561\ 509\ 916\ 7[$-$9]          &6.415\ 324\ 043\ 1[$-$12]          &5.011\ 809\ 33[$-$14]\\
50       &7.812\ 859\ 401\ 235[$-$10]          &2.630\ 602\ 571\ 9[$-$12]          &1.654\ 931\ 37[$-$14]\\
55       &4.210\ 472\ 655\ 409[$-$10]          &1.161\ 555\ 467\ 5[$-$12]          &5.999\ 556\ 2[$-$15]\\
60       &2.371\ 147\ 053\ 044[$-$10]          &5.443\ 579\ 080[$-$13]           &2.345\ 208\ 2[$-$15]\\
65       &1.383\ 617\ 655\ 412[$-$10]          &2.677\ 457\ 400[$-$13]           &9.748\ 095[$-$16]\\
70       &8.309\ 087\ 512\ 23[$-$11]           &1.369\ 821\ 733[$-$13]           &4.261\ 037[$-$16]\\
75       &5.106\ 469\ 950\ 92[$-$11]           &7.235\ 969\ 19[$-$14]            &1.940\ 914[$-$16]\\
80       &3.196\ 013\ 748\ 39[$-$11]           &3.921\ 694\ 89[$-$14]            &9.141\ 67[$-$17]\\
85       &2.028\ 253\ 121\ 49[$-$11]           &2.168\ 463\ 36[$-$14]            &4.421\ 83[$-$17]\\
90       &1.299\ 794\ 490\ 85[$-$11]           &1.216\ 900\ 77[$-$14]            &2.182\ 71[$-$17]\\
95       &8.376\ 878\ 675\ 0[$-$12]            &6.895\ 117\ 0[$-$15]             &1.092\ 81[$-$17]\\
100      &5.405\ 559\ 183\ 5[$-$12]            &3.923\ 335\ 2[$-$15]             &5.514\ 2[$-$18]\\
\end{tabular}
\end{ruledtabular}
\end{table*}

\begin{figure}[tbh]
\centering{
\includegraphics[width=8.4cm]{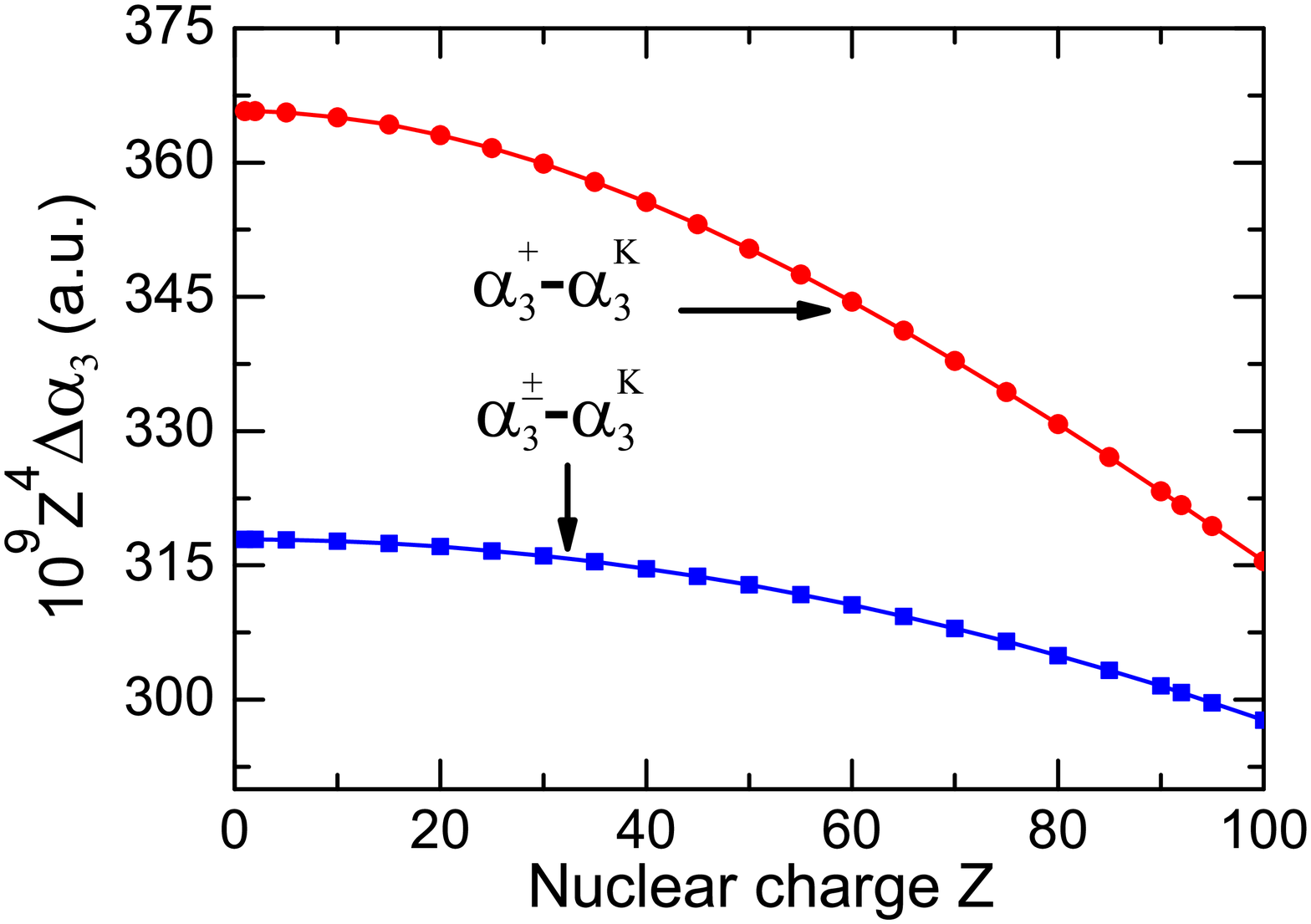}
} \caption{ \label{Fig5} (color online). Plot of $10^9 Z^4 \Delta \alpha_3$
as a function of nuclear charge, $Z$. }
\end{figure}

Figure~\ref{Fig5} plots $10^9 Z^4 \Delta \alpha_3$ as a function of
$Z$. These are seen to go to a constant value as $Z \to 0$. By an
analysis similar to that performed for the dipole polarizability, one
can deduce that $\alpha_3^{\pm}$, $\alpha_3^{+}$, and $\alpha_3^{\rm
K}$ are equal to order $O(\alpha^2 Z^2)$. The different $Z \to 0$
asymptotes for $\Delta \alpha_3^{\pm} $ and $\Delta \alpha_3^{+} $
indicate that $O(\alpha^4 Z^4)$ terms are different for
$\alpha_3^{\pm}$ and $\alpha_3^{+}$.

The analysis demonstrating that the differences between
$\alpha_2^{\pm}$ and $\alpha_2^{+}$ only appear at $O(\alpha^4 Z^4)$
has already been reported \cite{zhang12b}.   It has previously been shown
that these polarizabilities are agreement with the Kaneko $O(\alpha^2 Z^2)$
\cite{zhang12b}.   It is also
possible to plot $10^9Z^6 \Delta \alpha_4$ as a function of $Z$
giving plots similar to Figs.~\ref{Fig3} and \ref{Fig5}.
This demonstrates that $\alpha_4^{\pm}$ and $\alpha_4^{+}$ agree
with $\alpha_4^{\rm K}$ at the $O(\alpha^2 Z^2)$ level and the difference
between $\alpha_4^{\pm}$ and $\alpha_4^{+}$ occurs at the
$O(\alpha^4 Z^4)$ order.

\begin{table}
\caption{\label{S1tab} Comparison of the $S_1^+(0)$ sum rules. All digits
are stable with respect to further enlargement of the B-spline basis.  }
\begin{ruledtabular}
\begin{tabular}{lcc} \multicolumn{1}{c}{$Z$}&
\multicolumn{1}{c}{Present} &
\multicolumn{1}{c}{Ref.~\cite{drake81a}}
\\
\hline
1   & 0.999\ 955\ 631\ 350\ 807     & 0.999\ 9556 \bigstrut[t] \\
2   & 0.999\ 822\ 612\ 102\ 297     & \\
5   & 0.998\ 894\ 823\ 187\ 627     & \\
10  & 0.995\ 622\ 481\ 263\ 678     & 0.995\ 62\\
15  & 0.990\ 287\ 581\ 618\ 103     & \\
20  & 0.983\ 023\ 671\ 163\ 131     & 0.9830\\
25  & 0.973\ 973\ 703\ 862\ 452     & \\
30  & 0.963\ 278\ 628\ 607\ 378     & 0.9633\\
35  & 0.951\ 070\ 787\ 251\ 835     & \\
40  & 0.937\ 470\ 188\ 595\ 043     & 0.9375\\
45  & 0.922\ 582\ 481\ 520\ 977     & \\
50  & 0.906\ 497\ 887\ 620\ 449     & 0.9065\\
60  & 0.871\ 018\ 387\ 592\ 671     & 0.8710\\
70  & 0.831\ 424\ 017\ 561\ 149     & 0.8314\\
80  & 0.787\ 815\ 483\ 542\ 815     & 0.7878\\
90  & 0.739\ 933\ 345\ 752\ 064     & 0.7399\\
100 & 0.686\ 987\ 401\ 548\ 771     & 0.69\\
\end{tabular}
\end{ruledtabular}
\end{table}

\subsection{Sum-rules for the hydrogen isoelectronic series}

\begin{figure}[tbh]
\centering{
\includegraphics[width=8.4cm]{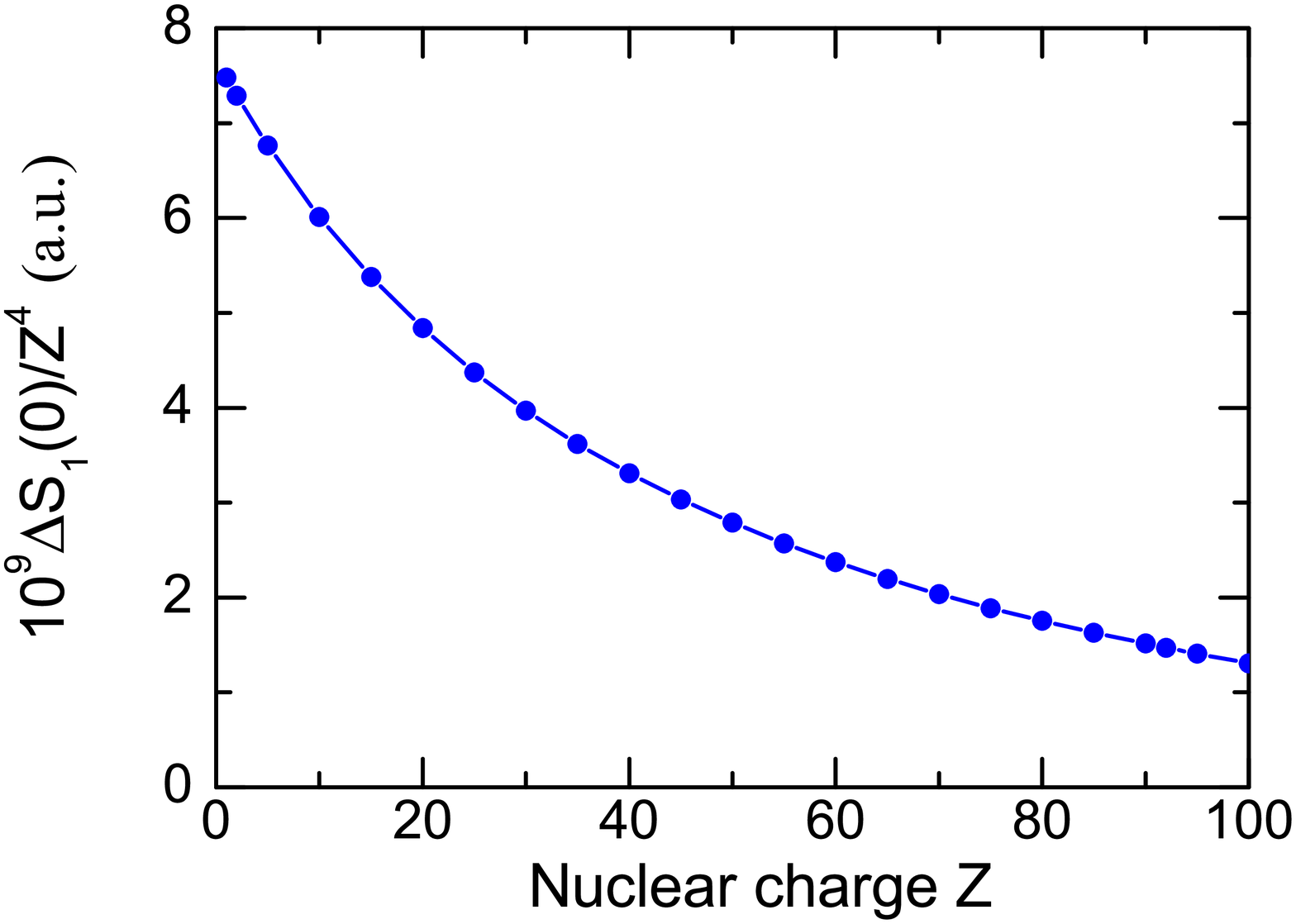}
} \caption{ \label{Fig6} (color online). Plots of $10^9 \Delta S_1(0)/Z^4$
versus nuclear charge, $Z$.  }
\end{figure}

The non-relativistic TRK sum-rule, $S_1(0)$ gives a value of unity
\cite{bell67a,lamm77b,inokuti71a}
for all hydrogen-like atoms and ions. However, $S^{\pm}_1(0)$
is exactly zero while the sum-rule, $S^{+}_1$(0) is almost equal to 1.  The appropriate
method to choose for the evaluation of the TRK sum rule has generated considerable discussion
\cite{levinger57a,drake81a,rustgi88a,aucar95a,romero98a,cohen03a,cohen04a,cohen05b,wullen12a}.
Table \ref{S1tab} compares the present B-spline values of $S^{+}_1(0)$
and compares them against the earlier calculation of Drake and Goldman
\cite{drake81a}. Keeping in mind the limited precision of the earlier
calculation, the agreement with the Drake and Goldman calculation
is perfect.

Figure \ref{Fig6} shows
\begin{equation}
\Delta S_1(0) = S^{+}_1(0)- S^{+ {\rm Levinger}}_1(0)    ,
\end{equation}
plotted as a function of $Z$.  It is noticed that $\Delta S_1(0)/Z^4$
goes to a constant as $Z \to 0$.  This demonstrates that the
present $S^{+}_{1}(0)$ is in agreement with the expression of Levinger
to order $O(\alpha^2Z^2)$.   It also demonstrates that the next
term in the expression for $S^{+}_1(0)$ occurs at the $O(\alpha^4 Z^4)$
level.  The near linear behaviour of $\Delta S_1(0)/Z^4$ at small
$Z$ indicates that the expansion for $S^{+}_1(0)$ contains a term of
$O(\alpha^5 Z^5)$.

While the B-spline calculations of $S^{+}(0)$ are compatible with
$O(\alpha^2Z^2)$ expressions \cite{levinger57a,cohen05b}, they cannot
be reconciled with the $O(\alpha^4Z^4)$ expression of Cohen.  A simple
analysis near $Z = 0$ suggests that
\begin{equation}
S^{+}_1(0) \approx 1 - \frac{5\alpha^2Z^2}{6} + 2.71  \ \alpha^4 Z^4 - 6 \alpha^5 Z^5 + O(\alpha^6Z^6)  \ .
\end{equation}
It has not been possible to reconcile the coefficient of 2.71  with Eq.~(8) of
Cohen \cite{cohen05b}.  But it is unclear how to interpret  $\langle p^4 \rangle$
of Eq.~(8) in \cite{cohen05b}.  The plot of $S^{+}_1(0)$ depicted in Fig.~3
of \cite{cohen05b} is certainly compatible with the present B-spline calculation.
However, Fig.~3 of \cite{cohen05b} plots the $O(\langle p^2 \rangle)$ approximation
to $S^{+}_1(0)$ and this is certainly not equal to $1 - \frac{5\alpha^2Z^2}{6}$.

\begin{figure}[tbh]
\centering{
\includegraphics[width=8.4cm]{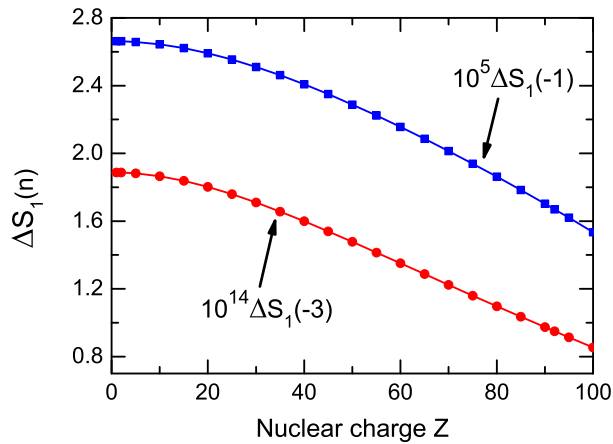}
} \caption{ \label{Fig7} (color online). Plots of $\Delta S_1(-1)$ and
$\Delta S_1(-3)$ versus nuclear charge, $Z$.  }
\end{figure}

Figure \ref{Fig7} shows the difference
\begin{equation}
\Delta S_1(n) = S^{\pm}_1(n)-S^{+}_1(n) \ ,
\end{equation}
plotted against $Z$ for $n = -1$ and $n = -3$.  It is noticed that
$\lim_{Z \to 0} \Delta S_1(-1)$ and $\Delta S_1(-3)$ both go to a
constant as $Z \to 0$.  Figure \ref{Fig3} established that
$\Delta S_1(-2)$ also has the same $Z \to 0$ limiting behaviour.
Writing either of the $S_1(n)$ in the form
\begin{equation}
S_1(n) = S^{\rm NR}_1(n) \left(1 + c_2 \alpha^2Z^2 + c_4 \alpha^4 Z^4 + \ldots \right)  \ ,
\end{equation}
allows one to deduce that the $c_2$ coefficients are different for
$S^{\pm}_1(-1)$ and $S^{+}_1(-1)$ since $S^{\rm NR}_1(-1) = 2/Z^2$.
However, one deduces that the $c_2$ and $c_4$ coefficients
are actually the same for $S^{\pm}_1(-3)$ and $S^{+}_1(-3)$ since
$S^{\rm NR}_1(-3) = 43/(4Z^6)$.

\subsection{Analytic expressions for the multipole polarizabilities of hydrogen-like ions}

\begin{figure}[tbh]
\centering{
\includegraphics[width=8.4cm]{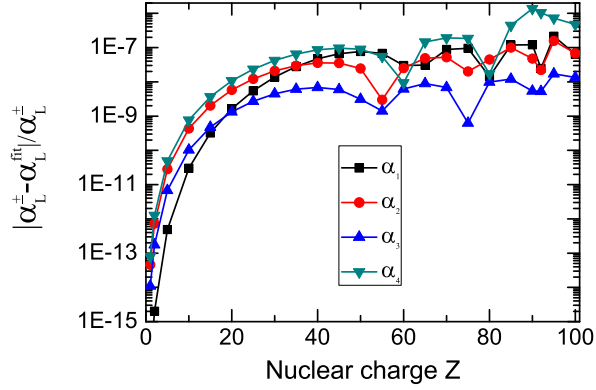}
} \caption{ \label{fit2} (color online). Relative difference between
$\alpha_{\ell}^{\pm}$ and the fits to this using Eq.~(\ref{powerseries})
as a function of nuclear charge, $Z$. }
\end{figure}

Analytic expressions were derived for $\alpha_{\ell}^{\pm}$ by
performing a least squares fit of the polarizabilities to
Eq.~(\ref{powerseries}).  The polarizabilities were divided by
the non-relativistic values prior to the fit.   The value of
$\lambda_{2}$ was fixed at
the values of Kaneko for $\ell = 1, 2$ and 3.  The $\lambda_2$
value for $\ell = 4$ was determined by evaluating Eq.~(36) of
Ref.~\cite{kaneko77a}.  The value of
$\lambda_4$ for $\alpha_1^{\pm}$ was set to the value from
Zon \cite{zon72a}.  Table \ref{fit3} lists the numerical values
of $\lambda_{2i}$ coming from the fit.  These coefficients give a
more precise representation of the exact dipole polarizabilities
than two previous representations \cite{drake81a,goldman89b}.  The
expressions for the quadrupole and octupole polarizabilities are
novel.

The quality of the fit to the B-spline $\alpha_{\ell}^{\pm}$ can be
seen from Fig.~\ref{fit2}.  The quality of the fits are of
very high accuracy at the smaller values of $Z$.  This occurs
since the leading $\lambda_2 \alpha^2 Z^2$ term uses the exact
value of $\lambda_2$.  The quality of the fit is degraded at
larger values of $Z$.  However, the maximum relative error in
the analytic expressions only exceeds one part per million
for values of $Z$ close to 100.


\begin{table*}
\caption{\label{fit3}
The Eq.~(\ref{powerseries}) fits to the multipole polarizabilities of
hydrogen isolectronic series ground states.  }
\begin{ruledtabular}
\begin{tabular}{lccccc}
\multicolumn{1}{c}{Term}&
\multicolumn{1}{c}{$\alpha_{1}^{\pm}$ \cite{goldman89b}} &
\multicolumn{1}{c}{$\alpha_{1}^{\pm}$} &
\multicolumn{1}{c}{$\alpha_{2}^{\pm}$} &
\multicolumn{1}{c}{$\alpha_{3}^{\pm}$} &
\multicolumn{1}{c}{$\alpha_{4}^{\pm}$} \\
[0.2ex]   \hline \\ [-1.5ex]
$\alpha^{\rm NR}$ &  $\displaystyle{\frac{9}{2Z^4}}$ & $\displaystyle{\frac{9}{2Z^4}}$   &  $\displaystyle{\frac{15}{Z^6}}$   &  $\displaystyle{\frac{525}{4Z^8}}$  &  $\displaystyle{\frac{8505}{4Z^{10}}}$ \bigstrut[t] \\ [0.6em]
$\lambda_2$      &   $\displaystyle{-\frac{28}{27}}$ & $\displaystyle{-\frac{28}{27}}$  &  $\displaystyle{-\frac{879}{600}}$  &  $\displaystyle{-\frac{5123}{2940}}$  & $\displaystyle{-\frac{33251}{17010}}$  \\  [0.5em]
$\lambda_4$      & 0.117\ 451\ 87(1)     & 0.117\ 451\ 870\ 668\ 402   &     0.502\ 471\ 315    &  0.854\ 144\ 263   & 1.177\ 235\ 432  \\
$\lambda_6$      & 0.007\ 482(1)       & 0.007\ 692\ 784      &  $-$0.014\ 151\ 521    &  $-$0.102\ 874\ 518 & $-$0.228\ 232\ 960  \\
$\lambda_8$      & 0.0010(1)         & $-$0.003\ 271\ 333     &     0.002\ 052\ 103      &  0.001\ 434\ 636    & 0.011\ 629\ 938  \\
$\lambda_{10}$   &    ---            & 0.006\ 117\ 861       &  $-$0.000\ 261\ 805     &  0.001\ 019\ 239    & $-$0.000\ 189\ 306   \\
$\lambda_{12}$   &    ---            &  $-$0.013\ 528\ 604      &    ---                &    ---            &  ---              \\ \hline
 &  \multicolumn{1}{c}{             } &
\multicolumn{1}{c}{$\alpha_{1}^{+}$} &
\multicolumn{1}{c}{$\alpha_{2}^{+}$} &
\multicolumn{1}{c}{$\alpha_{3}^{+}$} &
\multicolumn{1}{c}{$\alpha_{4}^{+}$}
\bigstrut[t] \\
[0.2ex] \hline \\ [-1.5ex]
$\lambda_2$      &                                 & $\displaystyle{-\frac{28}{27}}$  &  $\displaystyle{-\frac{879}{600}}$  &  $\displaystyle{-\frac{5123}{2940}}$  & $\displaystyle{-\frac{33251}{17010}}$ \bigstrut[t] \\  [0.5em]
$\lambda_4$     & ---      & 0.171\ 953\ 291        &     0.601\ 241\ 304    &  0.981\ 404\ 521   & 1.323\ 923\ 421  \\

$\lambda_6$   & ---        & $-$0.069\ 671\ 936     &  $-$0.164\ 544\ 588     &  $-$0.321\ 798\ 414 & $-$0.506\ 146\ 068  \\

$\lambda_8$    &   ---        & 0.075\ 248\ 612        &     0.118\ 942\ 451      &  0.171\ 613\ 014    & 0.236\ 007\ 349   \\

$\lambda_{10}$   &    ---            & $-$0.051\ 443\ 668     &  $-$0.057\ 448\ 561   &  0.073\ 735\ 494    & $-$0.093 122\ 365   \\
\end{tabular}
\end{ruledtabular}
\end{table*}

Equation (\ref{powerseries}) was also used to create an analytic expression
for $\alpha^{+}_{\ell}$.  In this case, the $\lambda_{2i}$ parameters with
$i > 1$ were treated as fitting parameters.  The results of the fit are
tabulated in Table \ref{fit3}.


\begin{table*}
\caption{\label{Cntab} The second-order dispersion coefficients (in
a.u.) for the H($1s$)-H($1s$) and H($1s$)-He$^+$($1s$) systems.
Results are given for the sum rules evaluated with and without the
states of the Dirac sea. All tabulated digits are accurate. The notation
$a[b]$ means $a \times 10^b$.}
\begin{ruledtabular}
\begin{tabular}{llll}
\multicolumn{1}{c}{ } &
\multicolumn{1}{c}{$C_n^{+}$} &
\multicolumn{1}{c}{$C_n^{\pm}$} &
\multicolumn{1}{c}{Non-relativistic} \\
\hline
\multicolumn{4}{c}{H($1s$)-H($1s$)} \bigstrut[t] \\
$C_{6}$     &    6.498\ 392\ 250\ 007\ 09     &   6.498\ 392\ 245\ 754\ 06      &  6.499\ 026\ 705\ 405\ 84    \\
$C_{8}$     &    1.243\ 840\ 307\ 694\ 35[2]  &   1.243\ 840\ 306\ 577\ 93[2]   &  1.243\ 990\ 835\ 836\ 22[2]   \\
$C_{10}$    &    3.285\ 370\ 791\ 861\ 60[3]  &   3.285\ 370\ 788\ 289\ 10[3]   &  3.285\ 828\ 414\ 967\ 42[3]   \\
\multicolumn{4}{c}{H($1s$)-He$^+$($1s$)} \\
$C_{6}$     &    0.657\ 548\ 755\ 759\ 311    &   0.657\ 548\ 758\ 416\ 787     &  0.657\ 716\ 656\ 238\ 770    \\
$C_{8}$     &    8.335\ 406\ 342\ 724\ 22     &   8.335\ 406\ 384\ 081\ 09      &  8.337\ 819\ 589\ 166\ 31   \\
$C_{10}$    &    1.588\ 773\ 716\ 759\ 79[2]  &   1.588\ 773\ 725\ 479\ 53[2]   &  1.589\ 267\ 575\ 526\ 71[2]   \\
\end{tabular}
\end{ruledtabular}
\end{table*}

\section{Dispersion coefficients}

The long-range dispersion interaction between two spherically symmetric
atoms can be written
\begin{equation}
V_{\rm disp}(R) \sim - \sum_{n=3}^{\infty}
\frac{C_{2n}}{R^{2n}}\ ,
\label{Vdisp1}
\end{equation}
The dispersion coefficients, $C_{2n}$ can be evaluated using
oscillator strength sum rules.  The explicit expression
is
\begin{eqnarray}
C_{2n}&=&\sum_{\ell_i=1}^{n-2} \sum_{\ell_j=1}^{n-2}
\delta_{n-1,\ell_i+\ell_j}
\frac{(2n-2)!}{4(2l_i)!(2l_j)!} \nonumber \\
&\times & \sum_{ij}
\frac {f_{A,gi}^{(\ell_i)}f_{B,gj}^{(\ell_j)}}{\varepsilon_{A,gi}\varepsilon_{B,gj}(\varepsilon_{A,gi}+\varepsilon_{B,gj})} ,
\label{Cn}
\end{eqnarray}
where $\ell_i + \ell_j +1 = n$ and
$\varepsilon_{A,gi}$ is the excitation energy from state $g$ to
state $i$ for atom $A$.
The sum implicitly includes the continuum, and
$f^{(\ell_i)}_{A,gi}$ is the oscillator strength of multipole $\ell_i$
connecting the state $g$ to the excited state $i$ for atom $A$.
Considerations of molecular symmetry do
not have a direct effect on Eq.~(\ref{Cn}) when both atoms are
in spherically symmetric states.

It is surprising that there has not yet been any calculation
of the hydrogen dimer dispersion coefficients based on oscillator
strengths from the Dirac equation.  This is rectified in Table~\ref{Cntab}
where the $C_{6}$, $C_{8}$ and $C_{10}$ coefficients are given for
two hydrogen atoms in their ground states.  Table~\ref{Cntab} also gives
the dispersion coefficients between a hydrogen atom and a He$^{+}$
ion.

The use of the Dirac equation leads to the H-H $C_6$ being reduced by
0.00063 a.u. or 0.0098$\%$.  The relative difference is about twice as
large as the difference between the relativistic and non-relativistic
polarizabilities.  The reduction in the size of $C_6$ is larger for
the H-He$^+$ system, being about 0.026$\%$.

\section{Conclusions}

A computational investigation based on B-spline methods has been
used to investigate the polarizabilities and related sum rules of the
hydrogen isoelectronic series.  Dipole polarizabilities have been
computed to a higher precision than any previous calculations.  One
distinction with previous calculations is that results were also reported
for calculations where the negative-energy Dirac sea is excluded
from the intermediate sum.  The
agreement with previously derived analytic expressions
\cite{zon72a,kaneko77a,yakhontov02a} for the
dipole polarizability could not be better.  High precision
calculations of the multipole polarizabilities for $\ell = 2, 3, 4$
are also given.
The present results provided a computational validation of
the earlier works of Kaneko \cite{kaneko77a} and Zon \cite{zon72a}.  The
$\alpha^{\pm}_{\ell}$ polarizabilities are in agreement with the Kaneko
expressions at the $O(\alpha^2 Z^2)$ level.  The $\alpha^{\pm}_{1}$
polarizability is also in agreement with the expressions of Zon \cite{zon72a}.
which includes terms at the $O(\alpha^4 Z^4)$ level.

Precise values for other oscillator strength sum rules have also
been computed.  The sum-rule $S^{\pm}_{\ell}(0) = \sum_i f^{(\ell)}_{gi} = 0$
provides a valuable check of the numerical reliability of the
calculations.  The sum-rule, $S^{+}_{1}(0)$ has been shown to be
compatible with the $O(\alpha^2Z^2)$ expression of Levinger {\em et al}
\cite{levinger57a}.  It is also compatible with earlier numerical calculations
\cite{drake81a,cohen05b}.

One aspect of the present work that represents a departure from earlier
work has been the treatment of the states of the negative energy sea.
Existing practice is that calculations of polarizabilities include
the states of the negative energy sea, while calculations of the
Bethe sum rule tend to omit these state.  The philosophy of the
present work has simply been to do two calculations for most
properties, those that include the states of the Dirac Sea and those
that omit them.

Analytic expressions for $\alpha^{\pm}_{\ell}$ and $\alpha^{+}_{\ell}$
to relative precisions not exceeding 10$^{-6}$ have been
obtained by fitting an $(\alpha Z)^n$ expansion to the computed
polarizabilities.  The $C_6$, $C_8$ and $C_{10}$ dispersion
coefficients for the long-range H-H and H-He$^+$ interactions were
also computed.

\begin{acknowledgments}

This work was supported by NNSF of China under Grant Nos.  11104323,
11034009 and by the National Basic Research Program of China under
Grant Nos. 2010CB832803 and 2012CB821305.  Jim Mitroy would like to thank the Wuhan
Institute of Physics and Mathematics for its hospitality during his
visits. The work of Jim Mitroy and Jun Jiang was supported in part by
the Australian Research Council Discovery Project DP-1092620.  We thank
Ting-Yun Shi and Scott Cohen for helpful discussions.
\end{acknowledgments}


\end{document}